\documentclass[10pt,journal,letterpaper]{IEEEtran}
\usepackage[letterpaper,top=0.7in,bottom=0.7in,left=0.65in,right=0.65in,columnsep=0.2in]{geometry}
\usepackage[T1]{fontenc}
\usepackage{newtxtext,newtxmath}
\usepackage{amsmath,graphicx,booktabs,cuted,capt-of,array,tabularx,longtable,calc,url,xurl}
\usepackage{enumitem}
\setlist[itemize]{leftmargin=1.15em,itemsep=1pt,topsep=3pt}
\setlist[enumerate]{leftmargin=1.3em,itemsep=2pt,topsep=3pt}
\usepackage{cite}
\usepackage[hidelinks]{hyperref}
\providecommand{\tightlist}{\setlength{\itemsep}{0pt}\setlength{\parskip}{0pt}}
\DeclareUnicodeCharacter{2212}{\ensuremath{-}}
\DeclareUnicodeCharacter{00D7}{\ensuremath{\times}}
\DeclareUnicodeCharacter{2264}{\ensuremath{\le}}
\DeclareUnicodeCharacter{2265}{\ensuremath{\ge}}
\DeclareUnicodeCharacter{03BB}{\ensuremath{\lambda}}
\DeclareUnicodeCharacter{0394}{\ensuremath{\Delta}}
\DeclareUnicodeCharacter{03B5}{\ensuremath{\varepsilon}}
\DeclareUnicodeCharacter{2192}{\ensuremath{\rightarrow}}
\DeclareUnicodeCharacter{2080}{\ensuremath{_0}}
\DeclareUnicodeCharacter{2081}{\ensuremath{_1}}
\DeclareUnicodeCharacter{2082}{\ensuremath{_2}}
\DeclareUnicodeCharacter{2083}{\ensuremath{_3}}
\DeclareUnicodeCharacter{2084}{\ensuremath{_4}}
\DeclareUnicodeCharacter{2085}{\ensuremath{_5}}
\hypersetup{pdftitle={Trigger Timing, Deadline Readiness, and Event-Aligned Accounting for Dynamic Ad Insertion},pdfauthor={Prashant Chaudhary and Kapil Khandelwal}}
\title{Trigger Timing, Deadline Readiness, and Event-Aligned Accounting for Dynamic Ad Insertion}
\author{Prashant Chaudhary and Kapil Khandelwal%
\thanks{(Corresponding author: Prashant Chaudhary.)}%
\thanks{P. Chaudhary is an independent researcher, Santa Clara, CA 95051 USA (e-mail: prashant@prashantchaudhary.com).}%
\thanks{K. Khandelwal is with the Department of Civil and Environmental Engineering and Earth Sciences, University of Notre Dame, Notre Dame, IN 46556 USA (e-mail: kapil.khandelwal@nd.edu).}}
\begin{document}
\maketitle
\begin{abstract}
An ad break in a live stream is a hard deadline. Once a cue arrives, the ad path has seconds to produce a state the player can use, and starting that work earlier buys time while giving the prepared state longer to expire before the break. Comparing trigger policies means scoring which of these happened; the records easiest to collect do not. We separate trigger, reach, readiness, playback, billability and measurement in an observed-event ledger, fix the reference-deadline rule before candidates are assigned, and account for contribution over every planned break--session pair, reached or not. Readiness means the declared state is valid at that deadline. We prove when scoring completion by the deadline instead reverses a policy comparison: for monotone playback with a shared latency law, a fixed state lifetime and common coefficients, the reversal occurs when the earlier policy's extra trigger cost falls between its validity gain and its completion gain. A synthetic study of 45,000 scripted sessions matches this condition on monotone paths, a check of its generator, and, outside its assumptions, finds the same reversal in mixtures with restarts, seeks, late joins and pauses. Playhead summaries misclassify up to 41\% of reach indicators yet reverse no comparison in the study: maximum-position errors cancel exactly, and terminal-position errors change only magnitudes. Completion scoring reverses the comparison in three of nine short-lifetime conditions, all at the one latency setting where the proved condition holds. The evidence is synthetic by design; code, seeds and event histories are public.
\end{abstract}
\section{Introduction}

Seconds before a live ad break, a cue reaches the stream. From that moment the ad decision service, the packager and the player share a fixed budget of time to produce something the viewer can watch when the break begins: selected ads, validated media and a player able to switch to them without a stall. Dynamic ad insertion (DAI) operates within that budget. Start the per-viewer work early and there is slack for auctions, retries and slow media; start it late and less work is spent on viewers who leave before the break. That sounds like a scheduling question. It is also a measurement question, because the records easiest to collect can favor the wrong policy.

Take two policies that differ only in when they start: EARLY begins 40 seconds before the break and LATE 10 seconds before. Preparation latency is lognormal with a median of 12 seconds (log-scale standard deviation 0.7), and a prepared state stays usable for 15 seconds, a deliberately short lifetime of the kind that prefetch windows and expiration settings make possible \cite{r8}. Under that law EARLY's job finishes in time with probability 96\% and LATE's with 40\%, so a comparison that counts finished preparation favors EARLY. At the moment the break starts, though, EARLY's state is still usable with probability only 10\%, against 40\% for LATE. Most of EARLY's finished states expire before the viewer needs them. With a lifetime this short, an early start trades deadline misses for expired states; Section IV-C states when scoring completion instead of validity reverses the comparison.

The example generalizes. Production telemetry makes it easy to merge events that a policy comparison has to keep apart: requests, playbacks, billable impressions, tracking beacons and independent measurements are different records \cite{r11,r12,r13,r14}, and a viewer who seeks past a break can leave the same terminal playhead as one who played through it. Each substitution yields arithmetic that adds up while counting the wrong events. The question also cuts across architectures. Server-side ad insertion (SSAI) and server-guided ad insertion (SGAI) can each resolve early or late \cite{r1,r2,r3,r4,r5,r6,r7,r8,r9,r10}; the timing questions studied here arise under either label, and this paper compares trigger policies, not architectures.

This paper makes three contributions. First, an event-timed model: an observed-event ledger, a reference deadline \(D_i\) whose rule is frozen before candidate assignment, readiness defined as validity of the declared state at \(D_i\), and contribution accounted over every planned pair (Sections III--IV). Second, two exact results on what shortcuts lose: Example~1 (Section III-B) shows that playhead summaries cannot in general recover trigger and reach once paths include restarts and seeks, and Proposition~1 (Section IV-C) gives the condition under which completion scoring reverses a comparison. Third, SIM1, a reproducible synthetic study of 45,000 scripted sessions shared by both policies, and an application protocol stating the joins, cost records and uncertainty treatment a deployment needs (Section V).

\textbf{Approach.} Every quantity is read off one ledger row per planned pair, so the contribution contrast never conditions on an outcome a candidate can change. Because the reference deadline is fixed before candidates are assigned, a candidate's own delays cannot move the target it is scored against, so readiness collapses to a single indicator, the validity of the declared state at \(D_i\). With readiness defined that way, the gap between completion and validity becomes an interval condition on preparation latency, and Proposition~1 follows by integrating that interval against the latency law. SIM1 then checks its generator against Proposition~1 where it applies, measures what happens where it does not (restarts, seeks, late joins, pauses and incomplete capture), and ships with code, seeds and event histories so that every SIM1 number can be regenerated.

Section II locates trigger timing in the standards and products that implement it. Sections III and IV define the events, the deadline and the accounting, ending with Proposition~1. Section V presents SIM1 and the application protocol, Section VI sets out scope and implications, and Section VII closes with open problems.

\section{Architecture, timing, and adjacent work}

An \textbf{ad opportunity}, or \textbf{avail}, is a planned point where ads or fallback may replace or interrupt primary content: a pre-roll precedes the content and a mid-roll interrupts it. An \textbf{ad pod} groups the components considered at one opportunity; fill is the share of those components that receive paid ads, and slate is filler shown when none is ready. A \textbf{manifest} tells the player which media to request. A \textbf{cue} announces an opportunity, while a \textbf{schedule} records expected opportunities independently; an opportunity is \textbf{actionable} when a permitted cue or schedule provides enough information to begin viewer-specific work. Makegoods and clawbacks are later credits or reversals of paid value, and invalid traffic is activity filtered out as nonhuman or fraudulent.

\subsection{Execution locus and resolution timing}

\textbf{Execution locus} identifies whether a server returns a personalized stitched presentation or the player resolves and switches to separately signaled content. \textbf{Resolution timing} identifies when personalized per-play work begins relative to the opportunity. The two are separate design dimensions, and a comparison that treats them as one ends up comparing labels. Under a stitched presentation, per-play work can begin at actionability or be deferred while assembly remains server-side. A player-guided path can preload decision resources or media, or wait until nearer playback. Earlier resolution allows more preparation time and exposes more nonreaching pairs to work; later resolution shortens both that lead and some speculative exposure. Together they determine scheduling, cache and load behavior.

Stitching refers to what the client receives as one presentation, whether or not the server concatenates files. Guided insertion does not require moving the auction into the client: selection may remain server-controlled while the player resolves, loads and executes a transition \cite{r1,r6,r7,r15}.

Timing also changes load. A synchronized live break can send many sessions to the ad decision service within seconds, and the resulting burst changes latency, fill and value; Amazon Web Services (AWS) documents retrieval windows and transactions-per-second shaping for this reason \cite{r8}. Trigger and load-control policies therefore have to be specified together.

\subsection{Standards and product mechanisms}

HTTP Live Streaming (HLS) Interstitials signal asset or asset-list resources through \texttt{EXT-\allowbreak X-\allowbreak DATERANGE} metadata \cite{r2}, \cite{r6}. The second-edition draft adds Preload Date Ranges for earlier resolution \cite{r2}. Dynamic Adaptive Streaming over HTTP (MPEG-DASH) edition 6 adds alternative Media Presentation Description (MPD) events for insertion or replacement \cite{r4,r5}. DASH Industry Forum (DASH-IF) Livesim2 documents replacement events, personalized List MPDs and \texttt{earliestResolutionTimeOffset} \cite{r3}, \cite{r33}. Earlier DASH-IF Part 5 supplies interoperability context \cite{r16}.

AWS Elemental MediaTailor shows how far two prefetch mechanisms can diverge \cite{r8,r9,r10,r17}. Schedule-based prefetching is listed as unsupported for SGAI \cite{r10}; guided prefetch instead depends on session-tagged HLS interstitial-media manifest refreshes, so caching \texttt{/v1/i-media} removes its trigger path, and the cited guide does not support it for DASH \cite{r9}. Equal schedule settings therefore need not imply equal request paths or costs.

A cue-driven path needs cue receipt and interpretation unless a declared independent schedule supplies actionability. Society of Cable Telecommunications Engineers (SCTE) standards 35-1 and 224 locate signaling and scheduling in that chain \cite{r27}, \cite{r28}. The Video Ad Serving Template (VAST) and Open Measurement supply metadata, tracking and measurement interfaces \cite{r11}, \cite{r12}, while OpenRTB's \texttt{tmax} bounds only the bid-response stage \cite{r18}. Auction and billable events remain distinct \cite{r14}. Interactive Advertising Bureau (IAB) and Media Rating Council (MRC) guidance separates loading from rendering and addresses indirect SSAI signals \cite{r29}, \cite{r30}. Table~\ref{tab:interfaces} maps these interfaces to the remaining evidence requirements.

\begin{table*}[!t]
\centering
\caption{Examples of documented interfaces and their measurement limits (interfaces checked on the access dates given in the references). This is a source-based interface map, not an instrumented vendor comparison.}
\label{tab:interfaces}
\fontsize{9}{10.5}\selectfont
\begin{tabular}{>{\raggedright\arraybackslash}p{0.2500\dimexpr\textwidth-6\tabcolsep\relax}>{\raggedright\arraybackslash}p{0.3750\dimexpr\textwidth-6\tabcolsep\relax}>{\raggedright\arraybackslash}p{0.3750\dimexpr\textwidth-6\tabcolsep\relax}}
\toprule
Implementation surface & Documented request, identifier, or event & What still needs measurement \\
\midrule
MediaTailor stitched workflow & Session initialization returns manifest and tracking URLs; client tracking retrieves avail/ad metadata \cite{r31}. & Player boundary crossing and the full declared playable-prefix criterion are not established by a successful metadata response. \\
MediaTailor guided workflow & Client-mode asset lists contain a \texttt{TRACKING} section rather than using \texttt{GetTracking}; guided prefetch uses session-tagged manifest refreshes \cite{r9}, \cite{r31}. & Heartbeat arrival is not proof that a particular viewer has reached the break or begun rendering. \\
Google server-guided DAI & Registration yields a stream ID and metadata/verification URLs; the client requests a pod manifest and sends activity pings \cite{r15}. & A pod response alone does not establish the common deadline state or contractual payment. \\
DASH-IF Livesim2 SGAI & The implementation documents session IDs, an ad decision endpoint, impression/quartile beacon collection, and session-status endpoints \cite{r33}. & Player transitions, production traces and calibrated viewer behavior. \\
\bottomrule
\end{tabular}
\end{table*}

\subsection{Relationship to adjacent work}

Predictive prefetching trades latency against speculative traffic \cite{r19}, and timing can change queueing even at equal transfer volume \cite{r20}. Real-time scheduling formalizes deadlines \cite{r21}. Prior DAI work addresses URL-based pointer mapping \cite{r22} and sixth-edition server-guided insertion \cite{r1}; adjacent cloud-caching work supplies retention policies \cite{r23}. Streaming studies relate playback problems and session characteristics to abandonment \cite{r24}, \cite{r25}, but supply no directly transferable DAI nonreach probability. What this paper adds is the DAI-specific event construction that joins the cue, session, pod, billability and reconciliation rules of Section II-B to observed-event accounting.

Two established results supply the method's tools. Comparing only viewers who reach an opportunity can compare different populations when the policy affects reach, which is the posttreatment-conditioning problem of causal inference \cite{r37}. The eligible-pair denominator and the frozen deadline are how a DAI comparison avoids that trap; identifying a causal effect still requires an appropriate assignment and observation scheme. We also rely on common random numbers \cite{r34} and missing-data bounds \cite{r36}.

The second result is that misclassification can distort a comparison and, under suitable error mechanisms, reverse its direction, as measurement-error research has shown, including Copeland et al.~\cite{r38} and the qualifications summarized by Yland et al.~\cite{r39}. Our contribution is to specify the DAI events and time conventions whose substitution produces such errors, and to state in Proposition~1 (Section IV-C) when one of them reverses a trigger-policy comparison.

\section{Event-timed model}

\subsection{The unit, the cohort, and the rules fixed in advance}

We analyze one planned \textbf{pod--play pair} (a break--session pair): a named mid-roll opportunity and an initiated session, retaining slot outcomes within the pod. A \textbf{candidate} comprises its implementation, trigger policy, load controls and routing rule. Label the actual routed path in a hybrid fleet. Fix eligibility, consent, device support, pod construction and the analysis horizon before assignment, retaining pairs that never reach the opportunity.

A \textbf{qualifying crossing} is the observed boundary crossing selected by a predeclared rule. Descriptive accounting may treat repeated crossings as separate pairs. A causal comparison instead retains candidate-induced seeks and restarts within the original pair or a separately declared session-level unit, without enlarging the cohort after assignment. Do not pool those denominators. For pre-roll, trigger and reach may both approach one while readiness still varies.

Index \(a\) identifies the candidate, \(i\) the planned pair, and \(t\) wall-clock time. A quantity indexed only by \(i\) is common to candidates for that pair. We suppress indices only when their meaning is unambiguous.

Declare one viewer-facing transition state that counts as ready for every candidate. A state is \textbf{admissible} only while it meets that criterion and remains valid and usable. A \textbf{dependency graph} comprises the processing steps and branches started under a trigger policy to create that state.

Four further rules are fixed before the comparison begins:

\begin{itemize}
\tightlist
\item
  the opportunity boundary is \textbf{frozen}, meaning that a candidate's own delays cannot move it;
\item
  the \textbf{session-expiry rule} says when a pair can no longer count as reached;
\item
  we record a \textbf{qualifying initiation} when a per-play resolution job starts and that start satisfies the declared trigger rule; and
\item
  the \textbf{primary-trigger rule} selects which qualifying initiation is used for timing and exposure accounting when more than one occurs.
\end{itemize}

The records fall into three groups.

\textbf{A reference clock shared by every candidate.} Two quantities are recorded once per pair, so no candidate can move the target it is scored against:

\begin{itemize}
\tightlist
\item
  \(B_i\): the target opportunity position on the declared primary-content timeline;
\item
  \(D_i\): the common wall-clock deadline at which the viewer-facing transition state is required.
\end{itemize}

\textbf{Trigger and reach events.} These record when work could start, when it did start, and whether the viewer got to the break. For each candidate and pair:

\begin{itemize}
\tightlist
\item
  \(A_{ai}\): the earliest wall-clock time at which the required cue or schedule is usable by candidate \(a\);
\item
  \(G_{ai}\): the scheduled media position at which candidate \(a\) intends a per-play trigger to occur, when candidate \(a\) defines such a position;
\item
  \(T_{ai}\): one when at least one qualifying initiation is observed, zero otherwise under complete trigger-event capture;
\item
  \(\tau_{ai}\): the wall-clock time selected by the primary-trigger rule, defined only when \(T_{ai}=1\);
\item
  \(R_{ai}\): one when the session reaches the frozen opportunity boundary before the common session-expiry rule applies, whatever is presented there, and zero otherwise.
\end{itemize}

\textbf{Attainment versus validity.} Producing the ready state once and holding it at the deadline are different events, and completion-based scoring fails in the gap between them. For each candidate and pair:

\begin{itemize}
\tightlist
\item
  \(Z_{ai}(t)\): a binary state indicator equal to one exactly while the common admissible state is valid and usable at wall-clock time \(t\), and zero otherwise;
\item
  \(H_{ai}\): one if a dependency graph started by a qualifying initiation ever attains the common state, and zero otherwise; and
\item
  \(C_{ai}\): the first wall-clock time at which that attainment occurs, defined only when \(H_{ai}=1\).
\end{itemize}

The definitions imply \(H_{ai}\le T_{ai}\), which completely captured records must satisfy. An externally prepared state may be deadline-valid without policy attainment (Section III-D). A missing required start or terminal record leaves the corresponding quantity unobserved, not zero.

Write \(T^*_{ai}\) for physical initiation and \(O^T_{ai}\) for adequate capture of that event class. Then \(T_{ai}=T^*_{ai}\) when \(O^T_{ai}=1\), and is unobserved otherwise; reach and validity follow the same construction. Sections III--IV assume complete capture; Section V-C4 relaxes it.

\textbf{The readiness window.} For a pair with \(T_{ai}=1\), the available wall-clock readiness window is

\[
W_{ai}=D_i-\tau_{ai}.
\tag{1}\label{eq:main-1}
\]

A primary trigger after the deadline gives \(W_{ai}<0\). On a cue-dependent or schedule-dependent path, \(\tau_{ai}\ge A_{ai}\), so the actionability-based upper bound is

\[
W^{\max}_{ai}=D_i-A_{ai},\qquad W_{ai}\le W^{\max}_{ai}.
\] If \(A_{ai}>D_i\), the cue-dependent path cannot meet the deadline. A separately prepared admissible state may still be valid at that deadline. Where \(G_{ai}\) is defined, the media distance \(B_i-G_{ai}\) is not a substitute for \(W_{ai}\): the two coincide only under further assumptions about continuous \(1\times\) playback, buffering, clock alignment, and the definition of the deadline.

The full notation and domains appear in Supplement S4.

For multiple qualifying initiations indexed by \(j\), record initiation times \(\tau_{aij}\). The default primary trigger is \(\tau_{ai}=\min_j\tau_{aij}\), with any alternative rule fixed before comparison. It opens the readiness window and trigger-cost exposure. Later retries and duplicates remain trigger-attributable costs without resetting \(\tau_{ai}\). Record a successful graph's own initiation separately when its processing time is needed.

Apply service-tier and avail-mapping eligibility before constructing pairs, and common session-expiry and background-playback rules afterward.

\textbf{Fixing the deadline.} We fix the rules that produce the opportunity boundary and deadline before candidate assignment. Their values for a particular pair need not be known before playback starts. For time-shifted live viewing, first map the authoritative opportunity presentation timestamp to a program-clock deadline \(D_i^{\mathrm{prog}}\). Let \(o_i^{(0)}\ge0\) be that session's validated live-edge delay, frozen from a measurement unaffected by candidate-specific work. Use the common reference deadline

\[
D_i=D_i^{\mathrm{prog}}+o_i^{(0)}.
\]

Share the frozen offset across candidates without re-estimating it from later playback. Record its measurement time, clock authority, event identifier, uncertainty, discontinuities, presentation-timestamp (PTS) restamping, and schedule update or cancellation rule. Cue arrival and interpretation set \(A_{ai}\) without moving \(D_i\). An unknown offset leaves viewer-relative readiness unresolved unless bounds permit classification. An origin-relative analysis may use \(D_i^{\mathrm{prog}}\) but must name that different target and keep its rate separate.

An offset measured before assignment, or after it when the recorded ordering shows that no candidate-specific work could have affected it, supplies the common target; one available only through candidate-dependent instrumentation cannot, and assigning earlier at device, household or channel-day level helps only if it also yields such an unaffected offset. Otherwise the analysis uses the separately named origin-relative target.

For on-demand playback, derive \(D_i\) from a candidate-independent reference trajectory to \(B_i\). A descriptive ledger may use a predeclared pause or seek rule. A causal comparison excludes post-assignment actions from that reference or treats them as candidate-affected outcomes. Applying the same update rule does not itself make deadlines common. Candidate stalls, retries and transition delays move neither \(B_i\) nor \(D_i\). Reach is one whether the boundary presents paid ads, fallback, slate, house content or primary-content continuation, independently of ad playback.

Readiness refers to the frozen reference schedule, including any validated candidate-unaffected offset. We report later drift, pauses, seeks and candidate-induced live-edge changes separately from readiness at that reference deadline.

\subsection{Observed trigger and reach events}

For any event \(\mathcal E\), let \(\mathbf 1\{\mathcal E\}\) equal one when the event occurs and zero otherwise. The two events at the center of the exposure accounting are

\[
\begin{aligned}
T_{ai}&=\mathbf 1\{\text{qualifying initiation observed}\},\\
R_{ai}&=\mathbf 1\{\text{boundary reached before session expiry}\}
\end{aligned}
\tag{2}\label{eq:main-2}
\]

Both are observed, and neither is inferred from a single terminal playhead value. A live join after the scheduled trigger opportunity, a catch-up trigger fired at join, a forward seek, a backward seek, a restart, and a skipped pod each break any rule that depends only on final media position.

\textbf{Example 1 (non-identification from playhead summaries).} Let the trigger position be 60 and the opportunity position be 100, and count a seek over the opportunity as a skip, not as reach. One session plays continuously from 0 to 110, restarts at 0, and stops at 10. A second plays from 0 to 40, seeks to 110, immediately restarts at 0, and stops at 10; assume no catch-up trigger. Both have final position 10 and maximum position 110. The first has \((T,R)=(1,1)\) and the second \((T,R)=(0,0)\). Thus neither the terminal position nor even the pair of terminal and maximum positions identifies these events over the allowed path class. The skip rule is explicit, and the point is structural: one counterexample suffices, however rare such paths are in a given service.

Write \(\Pr\) for probability under the declared eligible-pair population or synthetic script distribution. For a finite cohort, the corresponding empirical rates use its stated weights. Subscripts \(T\) and \(R\) on probabilities, and superscripts \(T\) and \(R\) on costs in Section IV-A, refer to the trigger and reach events; readiness-conditioning labels are defined in Section III-D, and Supplement S4 collects the notation. Define the trigger probability \(p_{T,a}=\Pr(T_{ai}=1)\) and the reach probability \(p_{R,a}=\Pr(R_{ai}=1)\). Where the conditioning event has positive probability, the reached-pair trigger rate \(g_a\) and the nonreached-pair trigger rate \(h_a\) are

\[
g_a=\Pr(T_{ai}=1\mid R_{ai}=1),
\qquad
h_a=\Pr(T_{ai}=1\mid R_{ai}=0).
\]

For \(0<p_{R,a}<1\), the law of total probability gives

\[
p_{T,a}=g_a p_{R,a}+h_a(1-p_{R,a}).
\tag{3}\label{eq:main-3}
\]

The same relationship can be written so that no conditional probability is ever taken over an empty cell. Let the first digit of a cell label record \(T_{ai}\) and the second record \(R_{ai}\), and define \(p_{10,a}=\Pr(T_{ai}=1,R_{ai}=0)\) and \(p_{01,a}=\Pr(T_{ai}=0,R_{ai}=1)\). Then

\[
p_{T,a}=p_{R,a}+p_{10,a}-p_{01,a}.
\tag{4}\label{eq:main-4}
\]

At \(p_{R,a}=0\), Equation~\eqref{eq:main-4} collapses to \(p_{T,a}=p_{10,a}=h_a\), and the empty reached-pair rate \(g_a\) drops out. At \(p_{R,a}=1\) it collapses to \(p_{T,a}=1-p_{01,a}=g_a\), and the empty nonreached-pair rate \(h_a\) drops out.

The \(p_{10,a}\) cell is the measure of trigger-without-reach exposure. How much of that exposure is waste depends on cancellation, which can cut the work short, and on reuse, whose output is allocated to the pairs it serves.

Because \(T_{ai}\) admits a late trigger as well as an early one, define the pre-deadline subset \(p^{\mathrm{pre}}_{10,a}=\Pr(T_{ai}=1,R_{ai}=0,\tau_{ai}<D_i)\) and report it wherever the timing distinction matters. The comparison \(\tau_{ai}<D_i\) is evaluated only on the \(T_{ai}=1\) subpopulation on which \(\tau_{ai}\) is defined. An initiation exactly at the deadline has \(W_{ai}=0\) and belongs to total triggering but not to this strictly pre-deadline subset. It does not replace the total trigger probability in the cost calculation of Section IV. Reported together, \(p_{10,a}\) and \(p^{\mathrm{pre}}_{10,a}\) distinguish total trigger-without-reach exposure from its pre-deadline subset.

Revenue earned by a reached pod stays attached to that pod and is not credited to a nonreaching pair merely because work was reused. Any contractually retained value, credit, or charge on an \(R_{ai}=0\) pair is handled as a separate nonreach branch in Section IV-A. The complementary cell \(p_{01,a}\) counts reached opportunities that lack the designated trigger, which includes missed-trigger cases and cases where some other permitted path had already prepared an admissible state.

An implementation that guarantees a designated per-play trigger on every reached pair satisfies \(R_{ai}\le T_{ai}\), which we call the \textbf{event-containment condition}. Under it, \(p_{01,a}=0\), and \(g_a=1\) whenever \(p_{R,a}>0\). For \(0<p_{R,a}<1\) this reduces Equation~\eqref{eq:main-3} to

\[
p_{T,a}=p_{R,a}+h_a(1-p_{R,a}).
\tag{5}\label{eq:main-5}
\]

The endpoint cases follow from Equation~\eqref{eq:main-4}; event containment additionally gives \(p_{T,a}=1\) when \(p_{R,a}=1\). Containment is a system property that complete event capture must confirm; placing a scheduled trigger before the opportunity does not establish it.

\subsection{Restricted monotone-playback representation}

Media position can stand in for the event ledger, but only inside a stratum where the playback assumptions hold. A \textbf{stratum} is a subset of pairs sharing the same trigger position, opportunity position, entry condition, and playback assumptions, indexed by \(s\); a conditioning suffix \(\mid s\) means the rate is computed within that stratum alone. Let \(G_s\) be the common trigger position and \(B_s\) the common opportunity position, with \(G_s\le B_s\). The representation additionally requires session entry at or before \(G_s\), a qualifying initiation when \(G_s\) is crossed, continuous forward playback, and no seeks, restarts, discontinuity jumps, or catch-up triggers. The observation stops at the common session-expiry boundary; a crossing recorded only after expiry is excluded. Expiry after a valid crossing does not revoke that earlier reach.

Let \(S_{ai}\) be the final media position reached before session expiry for pair \(i\) under candidate \(a\), and for any media-position threshold \(y\) define the stratum-specific survival function \(\overline F_{S,a\mid s}(y)=\Pr(S_{ai}\ge y\mid s)\). Then

\[
\begin{aligned}
T_{ai}&=\mathbf 1\{S_{ai}\ge G_s\}, & p_{T,a\mid s}&=\overline F_{S,a\mid s}(G_s),\\
R_{ai}&=\mathbf 1\{S_{ai}\ge B_s\}, & p_{R,a\mid s}&=\overline F_{S,a\mid s}(B_s),\\
h_{a\mid s}&=\frac{\overline F_{S,a\mid s}(G_s)-\overline F_{S,a\mid s}(B_s)}{1-\overline F_{S,a\mid s}(B_s)}.
\end{aligned}
\tag{6}\label{eq:main-6}
\]

Equation~\eqref{eq:main-6} assumes no particular abandonment distribution. The expression for \(h_{a\mid s}\) requires \(\Pr(S_{ai}<B_s\mid s)>0\), because the nonreached-pair trigger rate has no meaning when every session reaches \(B_s\). In that endpoint case, \(p_{R,a\mid s}=p_{T,a\mid s}=1\) and \(p_{10,a\mid s}=0\); \(h_{a\mid s}\) is not needed.

Strata do not pool freely. Combining strata with different opportunity positions, trigger rules, or included sessions does not generally return the average of the stratum-specific rates. When aggregating, weight the event-aligned value and variable-cost components, and apply shared horizon cost once to the pooled cohort unless a declared nonoverlapping rule allocates that cost across strata instead. For live and seekable services, estimate the event probabilities directly from joined logs, using Equation~\eqref{eq:main-4} when a conditioning cell is empty.

Reach stays candidate indexed throughout. Player scheduling, blocking work, incompatibility, and candidate-induced stalls can all affect pre-boundary behavior and abandonment. The boundary definition is common and frozen, but equal reach across candidates is a hypothesis to test, and nothing in the framework imposes an unconditional reach penalty on SSAI or on SGAI.

\subsection{Deadline readiness and downstream states}

Before comparison, declare a common viewer-facing criterion: required playable prefix, transition-timing tolerance and no remaining critical-path network action. Guided instructions with media still unfetched fail that criterion; server-validated stitched segments alone also do not prove client readiness. Each path needs evidence of the whole criterion. Supplement S7 instantiates it on two constructed paths.

Using the state variable \(Z_{ai}(t)\) from Section III-A, the deadline-readiness indicator is

\[
Q_{ai}=Z_{ai}(D_i).
\tag{7}\label{eq:main-7}
\]

A fallback that meets the common criterion at \(D_i\) has \(Q_{ai}=1\), and lower monetary value alone does not turn it into a miss. If no admissible state is valid at \(D_i\), then \(Q_{ai}=0\), and that holds whether the state arrives late or arrived early and has since expired, been evicted, or been revoked. Changing the readiness criterion changes the event definition, which invalidates an otherwise unchanged comparison.

\(Q_{ai}\) is readiness at the reference deadline, not at actual arrival. A post-assignment pause can delay arrival beyond \(D_i\), allowing a state valid at \(D_i\) to expire before it is needed. Arrival-relative readiness depends on a potentially candidate-affected time and cannot replace this candidate-invariant target. Section V-C3 reports it alongside reference readiness for the imposed, shared paths of the synthetic study (SIM1) in Section V-C.

A validity spell includes its attainment instant and excludes its expiry or invalidation instant. For one spell attaining the state at \(C_{ai}\) and ending at \(E_{ai}>C_{ai}\), where \(E_{ai}\) denotes that spell's invalidation time, the convention is \(C_{ai}\le t<E_{ai}\). Thus attainment exactly at \(D_i\) counts; expiry exactly at \(D_i\) does not. With multiple spells or prepared paths, \(Z_{ai}(t)\) is one when any admissible spell is active. A cancellation, failure or unresolved job is not ready unless another admissible state is active. Bounded validity is documented in MediaTailor: prefetch schedules define a consumption window, and recurring schedules can limit retrieved-ad availability by an expiration setting \cite{r8}. SIM1's 15-second lifetime is a stress value within that mechanism. A deadline probe observes one instant; clock uncertainty requires transitions across the uncertainty interval.

When \(T_{ai}=H_{ai}=1\) and \(\tau_{ai}\le C_{ai}\), define the policy trigger-to-attainment time, or preparation latency, \(L_{ai}=C_{ai}-\tau_{ai}\ge0\). It includes parallel or sequential work, retries, failed branches and speculative attempts.

A state predating \(\tau_{ai}\) or prepared outside the policy receives no \(L_{ai}\), and its source must be recorded. The inequality \(L_{ai}\le W_{ai}\) establishes first attainment by \(D_i\), equivalently \(C_{ai}\le D_i\), not continued validity. Expiry, eviction or revocation may still make \(Q_{ai}=0\). Logging can miss first attainment while directly observing readiness at \(D_i\). A stage timeout such as \texttt{tmax} affects attainment, fill or value without replacing either elapsed-time measure.

Timestamps must share a clock. Map \(A_{ai}\), \(\tau_{ai}\), \(C_{ai}\), the state-validity transitions, and \(D_i\) onto one, and let \(\varepsilon\ge0\) be a documented bound, in wall-clock units, on relative deadline-versus-state-transition alignment error. This is a bound on relative alignment, not a bound on each timestamp. Uncertainty intervals include both endpoints; they are not half-open validity spells. When the error bound is zero, the interval reduces to the single deadline. When only separate timestamp bounds are available, propagate both, as below. A pair is definitely ready only when \(Z_{ai}(t)=1\) throughout \([D_i-\varepsilon,D_i+\varepsilon]\), and definitely not ready only when the state is zero throughout that interval. Anything else is marked timing-indeterminate and carried into sensitivity analysis. Directional bounds are handled in Supplement S5.

Any reported difference built from these timestamps must propagate the bound of every timestamp inside it. Two symmetric per-timestamp bounds of size \(e\) give a worst-case difference bound of \(2e\); Supplement S5 gives the directional enclosure. An ordering whose difference interval crosses zero remains indeterminate. Do not drop indeterminate pairs: report lower and upper readiness rates by assigning those pairs zero and one respectively, or use a calibrated measurement-error model defined in advance.

\begin{figure*}[!t]
\centering
\includegraphics[width=7.16in]{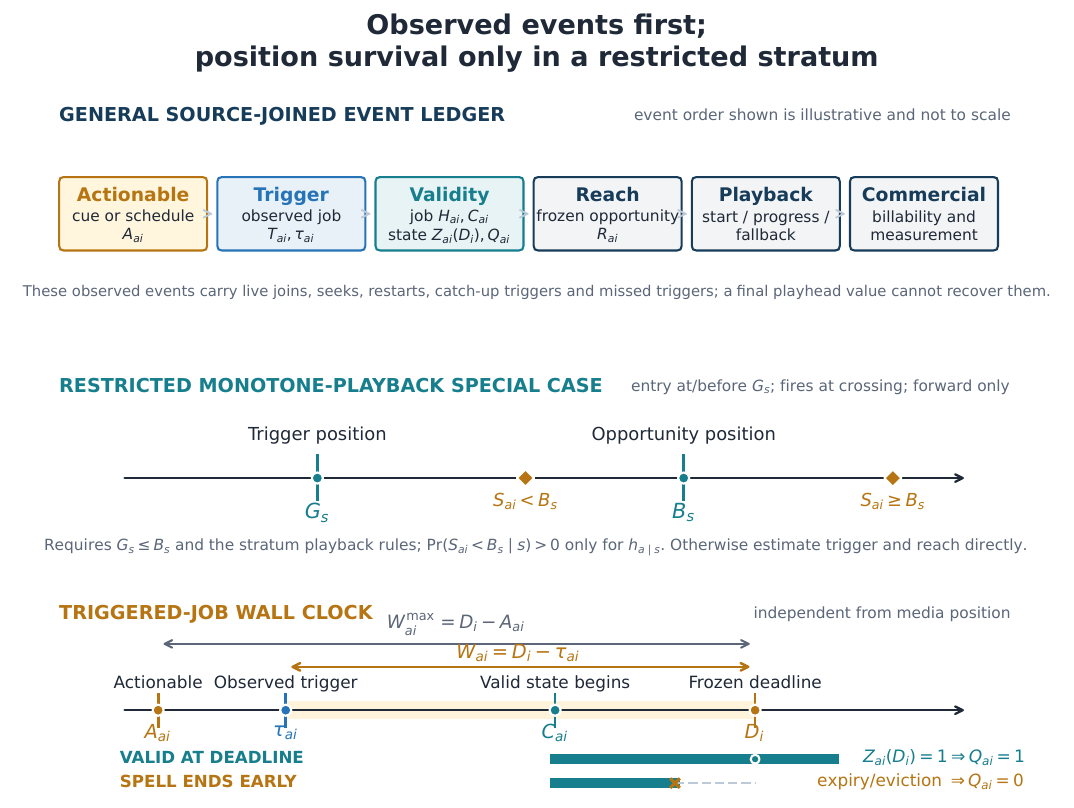}
\caption{The general model logs actionability, an observed trigger, state validity, frozen opportunity reach, playback, billability, and measurement. The wall-clock panel distinguishes a state valid at the deadline from one that expires or is evicted first. The media-position panel applies only to the continuous-forward-playback special case; live joins, seeks, restarts, and catch-up triggers require the observed event ledger.}
\label{fig:1}
\end{figure*}

Fig.~\ref{fig:1} collects the event ledger, the restricted position representation, and the wall-clock readiness relation defined above.

Three readiness rates distinguish conditioning and scoring time. Superscripts \(J\) and \(R\) condition reference-deadline readiness on a triggered job and a reached opportunity. Let \(t^{\mathrm{arr}}_{ai}\) be the first qualifying arrival time, defined only when \(R_{ai}=1\). Superscript \(\mathrm{arr}\) denotes scoring at arrival. On positive-probability conditioning events, define

\[
\begin{aligned}
q^J_a&=\Pr(Q_{ai}=1\mid T_{ai}=1),\quad
q^R_a=\Pr(Q_{ai}=1\mid R_{ai}=1),\\
q^{\mathrm{arr}}_a&=\Pr\!\left(Z_{ai}(t^{\mathrm{arr}}_{ai})=1\mid R_{ai}=1\right).
\end{aligned}
\tag{8}\label{eq:main-8}
\]

The job-conditioned rate includes cancellation and terminal errors, and can include readiness from shared or prepared paths outside the policy. Joint \((H_{ai},Q_{ai})\) cells distinguish policy attainment. If several paths supply \(Z_{ai}(D_i)\), record which supplied the deadline-valid state.

The reach-conditioned rate includes missed triggers and describes schedule-relative readiness among reached boundaries. All three rates are diagnostics, not extra multipliers in the event-weighted accounting of Section IV. Record actual arrival and post-departure cancellation separately.

A \textbf{hard readiness gate} rejects candidates below a specified readiness threshold. Use it only when a contract, service-level objective or nonnegotiable playback requirement supplies the threshold and target population. Otherwise incorporate misses and fallback into pod value. Estimate or bound the rates in Equation~\eqref{eq:main-8} directly without a parametric latency model, preserving viewer, session, stream, device, region and incident dependence in cluster or block resampling.

Illustrative path-specific readiness boundaries appear in Supplement S6.

\section{Pod-level economic accounting}

\subsection{Net value and event-aligned cost}

Let \(V^{\mathrm{net}}_{ai}\) be the complete pod's realized monetary value after reconciliation. The superscript denotes reconciliation, and the value may be positive, zero or negative. Record no-fill, partial fill, readiness miss, fallback, playback, billability, invalid-traffic filtering, discrepancy settlement, makegoods, credits, clawbacks and payment-dependent measurement conditions. Write \(\mathbb E\) for expectation.

Where \(p_{R,a}>0\), define \(v_a=\mathbb E[V^{\mathrm{net}}_{ai}\mid R_{ai}=1]\). Where the corresponding readiness cells also have positive probability, define \(v_a^{\mathrm{ready}}=\mathbb E[V^{\mathrm{net}}_{ai}\mid R_{ai}=1,Q_{ai}=1]\) and \(v_a^{\mathrm{miss}}=\mathbb E[V^{\mathrm{net}}_{ai}\mid R_{ai}=1,Q_{ai}=0]\). Readiness is then already inside \(v_a\):

\[
v_a=q^R_a v_a^{\mathrm{ready}}+(1-q^R_a)v_a^{\mathrm{miss}}.
\tag{9}\label{eq:main-9}
\]

Equation~\eqref{eq:main-9} uses occupied cells only. If \(q^R_a=1\) it reduces to \(v_a=v_a^{\mathrm{ready}}\) and the empty miss mean is omitted; if \(q^R_a=0\) it reduces to \(v_a=v_a^{\mathrm{miss}}\) and the empty ready mean is omitted.

The miss term is reconciled monetary pod proceeds, possibly zero or documented paid value usable after \(D_i\). Slate and primary-content continuation earn no pod revenue merely by preserving playback. Report engagement separately. An analysis starting from ready-cell value applies readiness once with explicit miss value, never again to the already averaged \(v_a\).

Conditional values belong to the candidate policy. Synchronized load and request timeouts can affect readiness, ready-state value or miss-state value, with the channel determined by measurement.

Let \(N>0\) be the number of planned pod--play pairs in the analysis cohort over the delivery horizon, and let \(z\) be the workload vector, whose components encode workload quantities or mix: channel count, opportunity count, device-class mix, distinct-creative count, creative-minutes, or reserved capacity.

Costs attach to events. Let \(\kappa^T_{ai}\ge0\) be the realized variable cost attributable to triggered work for candidate \(a\) and pair \(i\), covering downstream requests started by the trigger, redirects, retries, duplicate work that was not eliminated, and cleanup through the declared terminal state, cancellation point, or timeout cutoff. Let \(\kappa^R_{ai}\ge0\) be the realized downstream delivery, playback, reporting, and reconciliation cost attributable to a reached opportunity. Where the corresponding event has positive probability, the conditional cost means are

\[
c_{T,a}=\mathbb E[\kappa^T_{ai}\mid T_{ai}=1],
\qquad
c_{R,a}=\mathbb E[\kappa^R_{ai}\mid R_{ai}=1].
\]

Let \(K_a(N,z)\) be the total cost over the delivery horizon that neither variable-cost measure captures. It can include integration work, platform licenses, minimum commitments, capacity tiers, shared creative preparation, and effects on the primary manifest or content path.

Deduplication and allocation make the three cost buckets mutually exclusive. Keep retries, duplicate requests and cancellation cleanup in \(\kappa^T_{ai}\) unless explicitly allocated elsewhere, without also charging \(\kappa^R_{ai}\) or \(K_a(N,z)\).

Use unconditional event-weighted terms, assuming finite first moments and assigning zero to inactive branches. They remain defined at empty cells. When the relevant event probability is positive, the value, reach-cost and trigger-cost terms in Equation~\eqref{eq:main-10} below factor as \(p_{R,a}v_a\), \(p_{R,a}c_{R,a}\) and \(p_{T,a}c_{T,a}\).

The first of its three expectations gives the reached-pair value contribution. Using it as the entire value contribution assumes that reconciliation leaves no retained monetary value when \(R_{ai}=0\). Where the governing contract permits retained revenue, credits, or charges on a nonreached pair, replace it with the unconditional contribution \(\mathbb E[V^{\mathrm{net}}_{ai}]\) and report the \(R_{ai}=0\) branch separately.

Expected monetary contribution per planned pair is then

\[
\begin{aligned}m_a(N,z)={}&\mathbb E[R_{ai}V^{\mathrm{net}}_{ai}]-\mathbb E[R_{ai}\kappa^R_{ai}]\\&-\mathbb E[T_{ai}\kappa^T_{ai}]-\frac{K_a(N,z)}{N}.\end{aligned}
\tag{10}\label{eq:main-10}
\]

\textbf{Units and settlement.} All terms are currency per planned pair. The delivery horizon fixes \(N\) and cost allocation. Attribute later makegoods and clawbacks to the original cohort within a declared maturation window, without adding later traffic to \(N\). Report any settlement cutoff and remaining unsettled exposure.

\textbf{Reuse.} Book reuse once through its realized consequence. Documented avoided processing reduces incremental cost. Later-pod revenue stays in that pod's \(V^{\mathrm{net}}_{ai}\) and enters only a horizon containing that pair. Cross-cohort reuse needs common starting inventory and terminal adjustment, wash-in/wash-out, or another declared boundary rule for carried savings and contractually measurable terminal monetary value. Otherwise one cohort bears creation cost while another collects its benefit. Keep nonmonetary terminal value outside \(m_a(N,z)\) and \(c_{T,a}\).

\textbf{Shared-cost allocation.} Use incremental avoidable cost for architecture selection and report fully allocated cost separately for budgeting. Allocate shared costs by a causal driver, such as requests, creatives, creative-minutes, channels or capacity. Test disputed denominators that could reverse the comparison.

Let subscript \(0\) denote the current implementation, and compare \(\Delta m_{a,0}(N,z)=m_a(N,z)-m_0(N,z)\) with zero. The baseline does not itself have zero absolute contribution.

Compare candidates \(a\) and \(b\) at the same declared cohort size \(N\) and workload \(z\), and define the contribution difference as

\[
\Delta m_{a,b}(N,z)=m_a(N,z)-m_b(N,z).
\tag{11}\label{eq:main-11}
\]

\subsection{Cost regimes and pod expansion}

A fixed-cost assumption applies only within one implementation scope and capacity regime. Allocate commitments and tiers to \(K_a(N,z)\), and request-linked charges to \(\kappa^T_{ai}\) or \(\kappa^R_{ai}\) by their driving event. Evaluate tiers and shared costs at each candidate's workload.

Let \(\nu_{ai}\) count the realized required ad or fallback components. Because a count determined after triggering is an outcome, compare within \(\nu_{ai}\) groups only when the component count was fixed before assignment, and retain the count, component presence, pod outcome and fallback.

A zero-slot or no-fill result is not automatically ready; the readiness, value and fallback rules decide. Estimate readiness of all required components jointly rather than multiplying separate slot-level rates, and where one decision call resolves the whole pod, charge its shared decision cost once.

For sequential resolution, \(T_{ai}\) remains pair-level and \(\kappa^T_{ai}\) includes every initiation and attributable operation under the primary-trigger rule. When readiness requires all components, \(C_{ai}\) is their first joint attainment and \(L_{ai}\) the chain latency from primary initiation. Record component processing times separately. Sum slot values or costs only when interactions are immaterial, otherwise use the pod record.

Equation~\eqref{eq:main-10} gives the general comparison. Supplement S3 gives a local fixed-plus-per-pair break-even form, valid only at a fixed workload and in each candidate's cost regime.

\subsection{When completion scoring reverses the comparison}

Section III-D defines readiness as validity at the deadline. An easy shortcut scores completion instead: it counts a pair as ready when preparation finished by \(D_i\), whatever happened to the state afterward. Write \(\widetilde Q_{ai}=\mathbf 1\{H_{ai}=1,\ C_{ai}\le D_i\}\) for that proxy and \(\Delta\widetilde m_{a,b}(N,z)\) for the contrast of Equation~\eqref{eq:main-11} when each reached pod is valued as ready or missed according to \(\widetilde Q_{ai}\) instead of \(Q_{ai}\). The two contrasts can be compared exactly in the setting where the media position of Section III-C represents the events.

Consider candidates \(a\) and \(b\) whose pairs lie in monotone strata (Section III-C) with a common opportunity position \(B\), common playback paths and trigger positions \(G_a<G_b\le B\), and suppress the arguments \((N,z)\). Assume:
\begin{enumerate}[label=(\roman*)]
\tightlist
\item sessions enter at or before \(G_a\) and play at \(1\times\) on a program-aligned schedule whose deadline \(D_i\) is the wall-clock time at which continuous playback reaches \(B\), so every triggered pair has the fixed window \(W_a=D_i-\tau_{ai}=B-G_a\) or \(W_b=B-G_b\), with \(W_a>W_b\ge0\);
\item each trigger starts one job with no other prepared path; the job attains the common state at \(C_{ai}=\tau_{ai}+L_{ai}\), and the state stays valid on \([C_{ai},C_{ai}+\lambda)\) for a common lifetime \(\lambda>0\);
\item the latency \(L_{ai}\ge0\) has the same distribution function \(F\) under both candidates and is independent of the playback path;
\item playback, and therefore reach, is common to the two candidates, with \(p_R=\Pr(R_{ai}=1)>0\);
\item ready value, miss value, reach cost, trigger cost and shared cost are constants common to both candidates, with \(\Delta v=v^{\mathrm{ready}}-v^{\mathrm{miss}}\ge0\) and \(c_T\) the trigger cost.
\end{enumerate}
Write \(\pi=\Pr(G_a\le S_{ai}<G_b)\) for the extra trigger probability of the earlier candidate, and \(q_x=F(W_x)-F(W_x-\lambda)\) for \(x\in\{a,b\}\), with \(F(y)=0\) for \(y<0\).

\textbf{Proposition 1 (sign reversal under completion scoring).} Under (i)--(v),
\[
\Delta m_{a,b}=p_R\,\Delta v\,(q_a-q_b)-c_T\,\pi,
\tag{12}\label{eq:main-12}
\]
\[
\Delta\widetilde m_{a,b}=p_R\,\Delta v\,\bigl(F(W_a)-F(W_b)\bigr)-c_T\,\pi.
\tag{13}\label{eq:main-13}
\]
Completion scoring therefore reverses the comparison, \(\Delta m_{a,b}<0<\Delta\widetilde m_{a,b}\), exactly when
\[
p_R\,\Delta v\,(q_a-q_b)<c_T\,\pi<p_R\,\Delta v\,\bigl(F(W_a)-F(W_b)\bigr).
\tag{14}\label{eq:main-14}
\]
The opposite reversal cannot occur: \(\Delta\widetilde m_{a,b}-\Delta m_{a,b}=p_R\,\Delta v\,\bigl(F(W_a-\lambda)-F(W_b-\lambda)\bigr)\ge0\), so completion scoring can only move the comparison toward the earlier trigger. When \(\lambda>W_a\), the two contrasts are equal; they are also equal at \(\lambda=W_a\) when \(F(0)=0\).

\emph{Proof.} By (ii), a reached pair is ready exactly when \(C_{ai}\le D_i<C_{ai}+\lambda\), that is, when \(W_x-\lambda<L_{ai}\le W_x\); the proxy keeps only \(L_{ai}\le W_x\). Every reached pair is triggered, because sessions enter by \(G_a\) and \(G_x\le B\), so by (iii) and (iv) the readiness probabilities among reached pairs are \(q_x\) and \(F(W_x)\). By (iv) and (v), the reach-cost and shared-cost terms of Equation~\eqref{eq:main-10} are equal for the two candidates and cancel in Equation~\eqref{eq:main-11}, and the value terms differ by \(p_R\,\Delta v\) times the readiness difference. Trigger cost enters as \(c_T\,p_{T,x}\), and in the monotone stratum Equation~\eqref{eq:main-6}, applied to each candidate with the common survival function \(\overline F_{S}\) of the terminal position, gives \(p_{T,a}-p_{T,b}=\overline F_{S}(G_a)-\overline F_{S}(G_b)=\pi\). Equations~\eqref{eq:main-12} and \eqref{eq:main-13} follow, and \eqref{eq:main-14} restates their signs. Since \(W_a>W_b\) and \(F\) is nondecreasing, \(F(W_a-\lambda)\ge F(W_b-\lambda)\), and \(\Delta v\ge0\) makes the difference of the contrasts nonnegative. If \(\lambda>W_a\), both arguments are negative and the difference vanishes; at \(\lambda=W_a\) it equals \(p_R\,\Delta v\,F(0)\), which is zero when \(\Pr(L_{ai}=0)=0\). \(\square\)

Proposition~1 isolates one mechanism. An earlier trigger lengthens preparation time, and the same lead gives a finite-lifetime state time to expire; scoring completion credits the first effect and ignores the second. The bias, \(p_R\,\Delta v\,\bigl(F(W_a-\lambda)-F(W_b-\lambda)\bigr)\), values the excess of the earlier candidate's share of states that attain at least \(\lambda\) ahead of the deadline, and so expire by it, over the later candidate's share. When \(W_b<\lambda\), as in SIM1 (Section V-C1), the later candidate's states cannot expire before the deadline, the bias is \(p_R\,\Delta v\,F(W_a-\lambda)\), and earlier triggering or faster preparation raises it. Restarts, skips, late joins, pauses and catch-up triggers break assumption (i), and load feedback breaks (iii); SIM1's mixed-path conditions measure what happens then (Section V-C3).

\section{Measurement and application}

\subsection{Minimum event joins}

We join server and client evidence whenever the player owns part of the transition. Connected television (CTV) guidance distinguishes server records from observable client events \cite{r13}. Table~\ref{tab:observability} lists the evidence needed for each event class.

We retain a status, timestamp, source and join key for every event. An ephemeral session--opportunity identifier is sufficient. We check rendering through client evidence rather than server receipt, and account for cached or player-guided actions when a request is absent. We keep VAST tracking resources, Open Measurement results and contractual billable events distinct.

\begin{table*}[!t]
\centering
\caption{Minimum observability by event class.}
\label{tab:observability}
\fontsize{9}{10.5}\selectfont
\begin{tabular}{>{\raggedright\arraybackslash}p{0.2500\dimexpr\textwidth-6\tabcolsep\relax}>{\raggedright\arraybackslash}p{0.3750\dimexpr\textwidth-6\tabcolsep\relax}>{\raggedright\arraybackslash}p{0.3750\dimexpr\textwidth-6\tabcolsep\relax}}
\toprule
Event class & Server-side evidence & Client, joined, or external evidence \\
\midrule
Scope, cue, and trigger & Schedule, cue receipt, manifest, session, ad decision service (ADS), resolver, retry, duplicate, and cancellation timestamps. & Player receipt, actionability, primary-trigger rule, catch-up policy, session entry, shared event identifier, and clock alignment. \\
Opportunity reach and media path & Planned opportunity position and request proxies. & Frozen boundary event, actual presentation or continuation, playhead path, seeks, restarts, background state, skip, discontinuity, and session expiry. \\
Deadline and readiness & Configured target; ADS, validation, transcode, packaging, manifest, segment, expiry, eviction, and revocation timestamps. & Buffer, asset-list or MPD resolution, media loading, decode and validity transitions, common-clock error bound, transition policy, player/device version, and end-to-end trace. \\
Playback and fallback & Beacon receipt and server-side segment requests, with duplicate and cache caveats. & Ad start, quartiles, completion, stall, skip, slate, content continuation, and deduplication rules. \\
Billability and reconciliation & Ad-server, order, invalid-traffic, makegood, credit, and finance records. & Contract definition, qualifying playback event, settlement cutoff, and cohort attribution. \\
Independent measurement & Declared verification resources. & Supported measurement runtime, vendor evidence, coverage, and reconciliation; tracking and measurement remain distinct. \\
Cost & Cloud, content delivery network (CDN), platform, ADS, operations, and reserved-capacity records. & Delivery horizon, causal driver, allocation convention, tier, and sensitivity to disputed denominators. \\
\bottomrule
\end{tabular}
\end{table*}

MediaTailor also documents manifest, session-initialization and tracking-response log types, some requiring debug mode \cite{r32}. Common-state readiness still requires the joined evidence in Table~\ref{tab:observability}.

\subsection{Comparison design and data minimization}

Equation~\eqref{eq:main-10} is an accounting identity; a causal reading also needs the same target population and a defensible assignment rule. Prefer randomized session routing or cluster assignment. Where candidates share a constrained queue or service, the assignment cluster must include units that can affect one another through that load.

A non-serving shadow path compares only the outputs and load conditions it faithfully reproduces. An observational comparison must justify comparable groups, covariate overlap, adjustment for competing explanations and the target population. Disjoint groups support descriptive comparison; pairing them needs a justified construction \cite{r26}.

Assign candidates before candidate-specific actionability or triggering. If the candidate affects \(R_{ai}\), reach-conditioned diagnostics such as \(q_a^R\) compare post-assignment selected groups. They remain descriptive unless the analysis defends assumptions about pairs that would reach under both candidates \cite{r37}.

The target is the average contribution difference over all eligible planned pairs, reached or not, \(\Delta m_{a,b}(N,z)\) in Equation~\eqref{eq:main-11}, at the declared cohort size, workload, assignment unit and allocation regime. Under valid random assignment, correct measurement and control of shared-load effects, the unconditional event-weighted contrast, including \(K_a(N,z)/N\), estimates the policy effect on contribution. The same assumptions identify policy effects on unconditional components such as reach and the event-weighted trigger cost \(\mathbb E[T_{ai}\kappa^T_{ai}]\). Reach-conditioned summaries such as \(q^R_a\) and ready-cell values compare selected groups and remain diagnostics, and the decomposition does not attribute the effect to mechanisms. Individual routing on constrained shared infrastructure estimates a mixed-load policy, which may differ from an all-sessions deployment in latency, queueing, and the tier and reserved-capacity terms of \(K_a(N,z)\).

Specify trigger and concurrency controls together. Examine time, device, region, load and incident-window variation. A paired analysis needs both quantities on the same pair or another justified paired construction; shared aggregates alone do not establish pairing.

Collect only fields required by the event dictionary and strata. Where consent, technical, contractual or legal constraints prevent row-level joins, use compatible modular estimates or bounds. Propagate uncertainty and name unobserved quantities and coverage changes.

\subsection{Synthetic verification and stress demonstration (SIM1)}

Supplements S1--S2 work SIM1's monetary coefficients (listed below) through a stipulated example.

\subsubsection{Design and what it can test}

SIM1 checks that the analyzer computes the declared quantities and demonstrates the failure modes those definitions distinguish. The generator assigns a scripted path. The ledger and its bounds use only the declared event classes and capture metadata. The export also contains playhead records for the position comparators, but the ledger ignores them when an event class is missing. A separately implemented geometric oracle and event extractor check the computations.

We use a program-aligned synthetic \(1\times\) schedule in which position 100 corresponds to wall-clock deadline 100. EARLY and LATE initiate at positions 60 and 90. The clock-translation control adds 40 seconds to deadlines, session starts, actionability, session expiry and observation end. A late join at media position \(\eta\) starts at wall-clock \(\eta\) plus this translation.

The event rules are:
\begin{itemize}
\tightlist
\item \emph{Reach.} A continuous-play crossing counts as reach, a seek over the opportunity counts as a skip, and restarts remain within one pair.
\item \emph{Catch-up.} Both policies use catch-up on late-join scripts: entry at or beyond the policy's trigger position, including equality, starts the job. Skip scripts disable catch-up.
\item \emph{Jobs.} Each pair permits one job and no retriggering; jobs continue after departure, and their trigger cost covers that work.
\item \emph{Horizons.} Session expiry occurs at 180 seconds and observation ends at 400 seconds, with the same clock translation. Neither the expiry gate nor the actionability gate suppresses an event in the random study grid (below); constructed fixtures test the session-expiry gate, retries, cancellation, deduplication, missing records and prepared states (Supplement S8). The export omits invalidations after the observation end, so the loss procedure cannot delete them, and this convention determines which validity records can become unobserved.
\end{itemize}

The five path types and the three path mixtures M0--M2 appear in Table~\ref{tab:paths}. Restarts and skips require a seekable timeline (on-demand or time-shifted live), and late joins are live-shaped. M1 and M2 are stress geometries chosen to exercise every path type; they are not service profiles. The completion-versus-validity reversal also occurs in purely monotone M0, as Proposition~1 implies for these parameters, so it does not depend on seekability. Latency follows a lognormal distribution with median 2, 12 or 35 seconds, log-scale standard deviation 0.7 and a cap at 120 seconds. State lifetime is 15 or 300 seconds, and validity follows the half-open rule of Section III-D. Successive draws use the recorded pseudorandom stream.

\begin{table}[!t]
\centering
\caption{SIM1 path types and mixture weights. \(U[l,h]\) is a uniform draw; durations and positions are in seconds at unit playback rate, and every script quits after its last segment.}
\label{tab:paths}
\fontsize{8.5}{10}\selectfont
\begin{tabular}{@{}lccc>{\raggedright\arraybackslash}p{0.47\columnwidth}@{}}
\toprule
Path & M0 & M1 & M2 & Script \\
\midrule
Monotone & 1 & 0.60 & 0.20 & Play \(U[20,140]\) \\
Restart & 0 & 0.10 & 0.20 & Play \(U[105,120]\), restart at 0, play \(U[0,40]\) \\
Skip & 0 & 0.10 & 0.20 & Play \(U[20,50]\), seek to \(U[105,120]\), play \(U[0,20]\) \\
Late join & 0 & 0.10 & 0.20 & Enter at \(U[65,95]\), play \(U[0,50]\) \\
Pause & 0 & 0.10 & 0.20 & Play \(U[20,50]\), pause \(U[10,60]\), play \(U[20,70]\) \\
\bottomrule
\end{tabular}
\end{table}

The 18 base conditions (three path mixtures, three latency medians, two lifetimes) each contain five seeded replicates, each stored as one event shard, of 500 independent scripts (scripted sessions), giving 45,000 scripts. Within each script, the policies share the path and latency draw through common random numbers \cite{r34}. The clock copies give 36 represented conditions and 180,000 policy realizations. Trigger placement, opportunity position and coefficients remain fixed across this verification grid.

Terminal- and maximum-position comparators replace trigger and reach indicators while retaining actual deadline readiness. The completion comparator replaces only readiness, counting any attainment by the deadline. Every method uses hypothetical ready value 0.030, miss value 0.010, reached cost 0.003, trigger cost 0.002 and horizon cost per pair 0.002. We deliberately reuse Supplement S1's coefficients. Miss value represents a stipulated later paid outcome.

The incomplete-capture views delete 1\% or 5\% of the loss-susceptible records (initiation, reach and validity records) under nested masks. We draw masks separately for each policy and reuse each policy's mask in its translated copy. Common random numbers therefore couple paths and latencies; capture loss is drawn per policy. Capture metadata identify the affected event classes, and costs remain completely observed. The deleted records correspond to \(O=0\) in Section III-A; missing indicators remain unknown. Conservative bounds enumerate admissible reach/readiness states while retaining the observed costs. These are finite-sample feasible intervals rather than statistical confidence intervals \cite{r36}. A missing-as-zero comparator instead sets every unobserved \(T\), \(R\) or \(Q\) to zero in the constant-coefficient formula, which also drops the trigger cost of an unobserved initiation.

We use Wilson score intervals for proportions with positive denominators \cite{r35}. Contribution standard errors use paired script differences, and the artifact supplies all five replicate means. Neither clock copies nor alternative capture views enter the independent sample count. We report every condition without a significance filter.

\subsubsection{Event misclassification and cancellation in the policy contrast}

Both position shortcuts exactly reproduce trigger and reach in the monotone positive control M0. The nonmonotone mixtures expose two different errors: terminal position can forget an earlier crossing after a restart, while both terminal and maximum position can read a seek over an opportunity as a reach.

In M1, terminal-position reach misclassification ranges from 18.52\% to 20.40\% across the six latency/lifetime settings, which share one path law, so the spread is sampling variation; maximum-position misclassification ranges from 9.24\% to 10.44\%. The corresponding M2 ranges are 38.60--40.84\% and 18.48--21.16\%.

For M2 with median latency 12 seconds and lifetime 15 seconds, terminal position misclassifies 994 of 2,500 reach indicators: 39.76\%, with a 95\% Wilson score interval (reported only for a positive denominator) of 37.86--41.69\%. Nevertheless, its aggregate reach estimate is 41.64\%, compared with 42.68\% under the event ledger. False positives and false negatives partly offset one another: in Table~\ref{tab:paths} every restart path is a terminal-position false negative and every skip path a false positive for both shortcuts, so equal restart and skip weights make the terminal errors offset in expectation. Maximum position misclassifies 484 indicators, or 19.36\% (17.86--20.96\%), and reports 62.04\% reach. The EARLY nonreached-trigger rate \(h_a\) changes from 722/1,433 = 0.50384 under event accounting to 722/949 = 0.76080 under maximum position because the nonreached denominator is distorted.

\emph{Why the errors cancel.} Neither position shortcut reverses the EARLY-minus-LATE contribution contrast in any of the 18 base conditions. Maximum-position errors cancel exactly: on skip records both policies have \(T=R=Q=0\), while the shortcut assigns \(T=R=1\) and leaves \(Q=0\), so with policy-specific constant coefficients each false assignment adds \(v_a^{\mathrm{miss}}-c_{R,a}-c_{T,a}\), and the shared coefficients make that error equal on both sides in each realization, because common random numbers give both policies the same skip records.

\emph{What breaks the cancellation.} The path geometry matters: every skip in Table~\ref{tab:paths} starts before both trigger positions and lands past the opportunity, and a skip from between them would put the shortcut's false trigger on LATE alone, breaking the cancellation even with shared coefficients. Under the same marginal path law without paired draws, equality would hold only in expectation. Unequal policy-specific terms break the cancellation unless their net errors coincide; changing ready-cell value alone leaves these zero-readiness records untouched. Terminal position changes magnitudes without reversing signs. A policy difference can look right while its event levels are wrong.

\begin{table*}[!t]
\centering
\caption{Completion versus continued validity: the three sign-reversal conditions.}
\label{tab:reversals}
\fontsize{9}{10.5}\selectfont
\begin{tabular}{>{\raggedright\arraybackslash}p{0.1600\dimexpr\textwidth-10\tabcolsep\relax}>{\centering\arraybackslash}p{0.2200\dimexpr\textwidth-10\tabcolsep\relax}>{\centering\arraybackslash}p{0.2000\dimexpr\textwidth-10\tabcolsep\relax}>{\centering\arraybackslash}p{0.2200\dimexpr\textwidth-10\tabcolsep\relax}>{\centering\arraybackslash}p{0.2000\dimexpr\textwidth-10\tabcolsep\relax}}
\toprule
Path mixture & Event-ledger contrast & MC standard error & Completion-proxy contrast & MC standard error \\
\midrule
M0 & \(-0.0025296\) & 0.0001555 & \(+0.0031104\) & 0.0001600 \\
M1 & \(-0.0023328\) & 0.0001647 & \(+0.0034592\) & 0.0001637 \\
M2 & \(-0.0016744\) & 0.0001673 & \(+0.0033816\) & 0.0001611 \\
\bottomrule
\end{tabular}
\par\smallskip\begin{minipage}{\textwidth}\footnotesize All entries are hypothetical currency units per planned pair; MC denotes Monte Carlo. Each row uses median latency 12 seconds and usable-state lifetime 15 seconds.\end{minipage}
\end{table*}

\subsubsection{Completion and deadline validity}

Here the contribution contrast is \(\Delta m_{\mathrm{EARLY},\mathrm{LATE}}\) from Equation~\eqref{eq:main-11}, and readiness among reached pairs is \(q_a^R\). We chose the short lifetime to expose expiry as a stress mechanism. Proposition~1 gives in closed form what M0 should show in expectation. For a reached monotone script the EARLY and LATE windows are 40 and 10 seconds, so at a 15-second lifetime EARLY is ready only when its latency lies in \((25,40]\) and LATE when it lies in \([0,10]\), while the completion proxy accepts \([0,40]\) and \([0,10]\). This interval form uses the one-spell model; the general ledger admits multiple spells. The discrepancy is designed in; SIM1 measures its consequences, carries it into mixed paths, and checks the arithmetic.

The completion comparator reverses the comparison in three of the nine short-lifetime conditions. All have a 12-second median latency and 15-second lifetime, with one reversal at each path-mixture level. The clock-translated copies reproduce these outcomes exactly; they do not constitute additional independent reversals. Table~\ref{tab:reversals} gives the three contrasts and their Monte Carlo standard errors.

Proposition~1 explains the M0 pattern and locates its boundary. At a 2-second median almost every job completes within LATE's 10-second window; completion scoring slightly favors EARLY, but in M0 that value gain is smaller than EARLY's extra trigger cost. At 12 seconds the latency law gives EARLY a completion probability of 0.9573 but a validity probability of only 0.1045, against 0.3973 for LATE under both definitions, and Equation~\eqref{eq:main-14} holds. At 35 seconds both definitions favor EARLY. With M0's \(p_R=1/3\), \(\pi=1/4\), these coefficients and the 15-second lifetime, Equation~\eqref{eq:main-14} holds for medians between about 3.7 and 20.9 seconds, and again between about 97.5 and 109.4 seconds (the M0 reversal bands), where few jobs finish in time under either policy and both contrasts stay below 0.0002 in magnitude. The 12-second setting is the only grid point inside either band; the M0 reversal sits there, and the M1 and M2 reversals occur at the same setting. Equations~\eqref{eq:main-12} and \eqref{eq:main-13} reproduce both contrasts in all six M0 conditions within 1.2 Monte Carlo standard errors, for example \(-0.00245\) and \(+0.00323\) at 12 seconds against the estimates \(-0.00253\) and \(+0.00311\) in Table~\ref{tab:reversals}. Path-specific exposure and the shared coefficients determine the mixed-path contrasts, which lie outside Proposition~1.

\begin{figure*}[!t]
\centering
\includegraphics[width=7.16in]{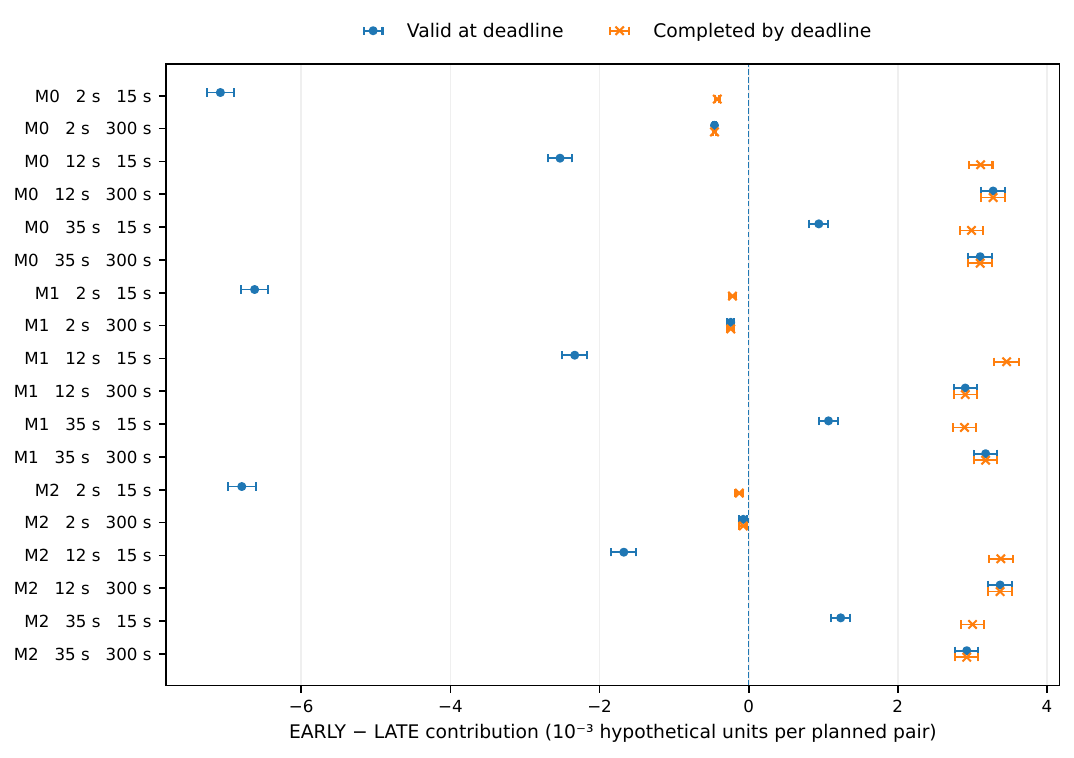}
\caption{EARLY-minus-LATE contribution in every base condition, using the event ledger and completion proxy. Labels identify mixture, median latency and state lifetime in seconds. The horizontal scale is in \(10^{-3}\) hypothetical currency units per planned pair. Horizontal error bars show one Monte Carlo standard error of each paired-script contrast, not a 95\% confidence interval. Table~\ref{tab:reversals} and the artifact give the underlying uncertainty and replicate results. Neither the translated copies nor masks add independent samples.}
\label{fig:2}
\end{figure*}

Fig.~\ref{fig:2} and Table~\ref{tab:conditions} report every base condition, and unrounded values accompany the code artifact. In the M2 condition with median latency 12 seconds and lifetime 15 seconds, EARLY readiness among reached opportunities is 216/1,067 = 20.24\% when current validity is required, but 848/1,067 = 79.48\% when any completion before the deadline is counted. LATE readiness is 376/1,067 = 35.24\% under both definitions in that condition.

A restart trace from that condition, the first restart script in its first shard, illustrates the distinction. The viewer crosses the opportunity at time 100, continues to media position 112.3198, restarts and ends at 28.2250. The EARLY job becomes valid at 63.7355 and expires at 78.7355. The LATE job becomes valid at 93.7355 and expires at 108.7355. At the frozen deadline of 100, both policies have reached the opportunity, but only LATE has a usable state. Their configured contributions are 0.003 and 0.023. Terminal position marks both pairs unreached, while completion-by-deadline marks both ready. Each shortcut erases the pair's \(-0.020\) contribution contrast, each for its own reason.

The nine 300-second-lifetime conditions are analytic consistency controls: any attained state starts no earlier than the translated time origin, and the deadline is 100 seconds after that origin. A state completed by the deadline therefore cannot have expired there. Completion-proxy disagreement, and hence a reversal caused by that proxy, is impossible in these nine conditions; for monotone paths this is the case \(\lambda>W_a\) of Proposition~1. M0 playhead agreement is likewise guaranteed by its monotone construction.

Scoring the same reached pairs at actual arrival rather than at the reference deadline changes EARLY readiness by \(-1.36\) to \(+4.51\) percentage points and LATE readiness by \(+0.18\) to \(+5.54\) percentage points across the 12 M1/M2 conditions. Every changed indicator lies on a pause path: all 11,314 nonpause reached scripts are invariant under both scoring times. The mechanism is forced by the design. Each pause precedes both trigger positions and exceeds LATE's 10-second reference window, so LATE is never ready at the reference deadline on a pause path and recovers its full window at arrival, while EARLY gains readiness through later attainment or loses it through expiry. Rescoring the all-pair contributions at arrival with unchanged coefficients keeps all 12 signs (Supplement S9).

The arrival comparison is a sensitivity reading: arrival time can itself be candidate-affected (Section III-D), and the comparison is paired here only because SIM1 imposes paths and arrivals. The 300-second rows, whose readiness can change between deadline and arrival, keep their control role: the controls compare completion with validity at one fixed deadline, and work unfinished at that deadline can finish by a later arrival, which is a different comparison.

\begin{table*}[!t]
\centering
\caption{EARLY-minus-LATE contribution across all 18 base conditions.}\label{tab:conditions}
\fontsize{9}{10.5}\selectfont
\begin{tabular}{>{\raggedright\arraybackslash}p{0.1000\dimexpr\textwidth-14\tabcolsep\relax}>{\centering\arraybackslash}p{0.1000\dimexpr\textwidth-14\tabcolsep\relax}>{\centering\arraybackslash}p{0.1000\dimexpr\textwidth-14\tabcolsep\relax}>{\centering\arraybackslash}p{0.1750\dimexpr\textwidth-14\tabcolsep\relax}>{\centering\arraybackslash}p{0.1750\dimexpr\textwidth-14\tabcolsep\relax}>{\centering\arraybackslash}p{0.1750\dimexpr\textwidth-14\tabcolsep\relax}>{\centering\arraybackslash}p{0.1650\dimexpr\textwidth-14\tabcolsep\relax}}
\toprule
Mix & Median (s) & Lifetime (s) & Ledger & Terminal & Maximum & Completion \\
\midrule
M0 & 2 & 15 & \(-0.0070848\) & \(-0.0070848\) & \(-0.0070848\) & \(-0.0004208\) \\
M0 & 2 & 300 & \(-0.0004584\) & \(-0.0004584\) & \(-0.0004584\) & \(-0.0004584\) \\
M0 & 12 & 15 & \(-0.0025296\) & \(-0.0025296\) & \(-0.0025296\) & \(+0.0031104\) \\
M0 & 12 & 300 & \(+0.0032784\) & \(+0.0032784\) & \(+0.0032784\) & \(+0.0032784\) \\
M0 & 35 & 15 & \(+0.0009408\) & \(+0.0009408\) & \(+0.0009408\) & \(+0.0029888\) \\
M0 & 35 & 300 & \(+0.0031048\) & \(+0.0031048\) & \(+0.0031048\) & \(+0.0031048\) \\
M1 & 2 & 15 & \(-0.0066256\) & \(-0.0047216\) & \(-0.0066256\) & \(-0.0002176\) \\
M1 & 2 & 300 & \(-0.0002400\) & \(-0.0002720\) & \(-0.0002400\) & \(-0.0002400\) \\
M1 & 12 & 15 & \(-0.0023328\) & \(-0.0017088\) & \(-0.0023328\) & \(+0.0034592\) \\
M1 & 12 & 300 & \(+0.0029048\) & \(+0.0018568\) & \(+0.0029048\) & \(+0.0029048\) \\
M1 & 35 & 15 & \(+0.0010704\) & \(+0.0005824\) & \(+0.0010704\) & \(+0.0028944\) \\
M1 & 35 & 300 & \(+0.0031776\) & \(+0.0021936\) & \(+0.0031776\) & \(+0.0031776\) \\
M2 & 2 & 15 & \(-0.0067976\) & \(-0.0028936\) & \(-0.0067976\) & \(-0.0001256\) \\
M2 & 2 & 300 & \(-0.0000720\) & \(-0.0001040\) & \(-0.0000720\) & \(-0.0000720\) \\
M2 & 12 & 15 & \(-0.0016744\) & \(-0.0004344\) & \(-0.0016744\) & \(+0.0033816\) \\
M2 & 12 & 300 & \(+0.0033720\) & \(+0.0011240\) & \(+0.0033720\) & \(+0.0033720\) \\
M2 & 35 & 15 & \(+0.0012344\) & \(+0.0003624\) & \(+0.0012344\) & \(+0.0030024\) \\
M2 & 35 & 300 & \(+0.0029248\) & \(+0.0007968\) & \(+0.0029248\) & \(+0.0029248\) \\
\bottomrule
\end{tabular}
\par\smallskip\footnotesize Seven decimal places are shown for comparison with the artifact outputs rounded to the same precision; the artifact retains additional digits. This is display precision: Monte Carlo standard errors for the contrasts in this table range from 0.0000252 to 0.0001868 (Fig.~\ref{fig:2}, Table~\ref{tab:reversals} and the artifact).
\end{table*}

\subsubsection{Incomplete telemetry and near-zero comparisons}

Bounds use event-class records and capture metadata, not event occurrence inferred from retained costs, even where the cost schedule implies it. They are therefore valid conservative feasible intervals and can be wider than the sharp identification region given all retained variables.

At 1\% susceptible-event loss, unknown deadline readiness ranges from 0.32\% to 1.84\% across policy-condition combinations; at 5\% loss it ranges from 1.92\% to 8.00\%. The ledger preserves this uncertainty rather than filling missing event classes from exported playhead records or generator internals.

Conservative contribution-contrast bounds include zero in one of the 18 base conditions at 1\% loss and six at 5\% loss. For M2, median latency 12 seconds and lifetime 15 seconds, the 5\% bounds are {[}\(-0.0025116\), \(-0.0004684\){]}, preserving a negative contrast in this finite generated sample. For M2, median latency 2 seconds and lifetime 300 seconds, they are {[}\(-0.0013172\), +0.0011412{]}. Even with complete capture, the latter condition is close to zero: its mean contrast is \(-0.0000720\), Monte Carlo standard error 0.0000549, and one replicate mean is positive. It supports no policy preference. No missing-as-zero point comparison reverses sign in this grid.

Correlated record loss could move these bounds in either direction relative to the independent-loss realization reported here. The guaranteed ordering is narrower: for nested information sets under the same constraints, removing information cannot shrink the feasible set. How fragile a field comparison is depends on field loss patterns, which these masks omit.

\subsubsection{Verification and scope}

Regeneration from the recorded seeds matched all 180,000 realizations and 540,000 observation views: complete capture, 1\% loss and 5\% loss for each realization. A separate event extractor and integer-milliunit checker reproduced primary quantities and paired summaries. Monotone agreement, occupied-cell identities, conservative bounds and clock-translation invariance passed. Latencies use correctly rounded exponential and logarithm functions, so regeneration gives identical records on Linux x86-64 and macOS arm64 (Supplement S10).

The post-study arrival diagnostic reads retained records without new draws or generator execution; its separately implemented checker matched all 120,000 policy/clock classifications and all 72 aggregate rows.

The geometric oracle uses the generated script and parameters, while the separate event extractor uses the exported events and an integer monetary calculation. Their trigger, reach and readiness computations are distinct from the analyzer's; verification orchestration reuses the generator and masking utilities. The arrival checker shares recorded inputs and declared semantics with its scorer, but reconstructs crossings and validity separately and imports neither the scorer nor simulator. The M0 contrasts agree with Equations~\eqref{eq:main-12} and \eqref{eq:main-13} within Monte Carlo error (Section V-C3), a closed-form check on the monotone path and latency draws; the other path types have no such check of the common sampler, so a stable generation error there could reproduce itself.

\subsection{Application-protocol specification}

Six steps specify how an analyst applies the method.

\begin{enumerate}
\def\labelenumi{\arabic{enumi}.}
\tightlist
\item
  \textbf{Define the comparison before assignment.} Name the eligible cohort, actual routed implementation, trigger/load policy, opportunity, delivery horizon and settlement lag. For deployment selection, include the current implementation; SIM1 instead compares two hypothetical candidates.
\item
  \textbf{Freeze the event dictionary.} Fix crossing, trigger, readiness, deadline, expiry, fallback and reconciliation rules. Keep playback, billability and measurement separate.
\item
  \textbf{Join the evidence.} Use per-pair keys and source/clock metadata. Report \(W_{ai}\) only for triggered pairs; omit empty-cell conditionals and bound unobserved states instead of zero-filling them.
\item
  \textbf{Apply justified requirements.} Protocol, privacy and device requirements may constrain admission. A readiness threshold needs external operational or contractual grounds.
\item
  \textbf{Compute contribution.} Attribute realized value and mutually exclusive costs with Equation~\eqref{eq:main-10}; allocate reuse and horizon costs once. Keep unsettled exposure explicit.
\item
  \textbf{Interpret with the design.} Use joint uncertainty only where pairing or assignment supports it, and separate conditional diagnostics from causal effects. Report indeterminate when supported comparison intervals cross zero.
\end{enumerate}

The event-dictionary template in the artifact supplies editable fields for these steps and a constructed example row.

\section{Implications and limitations}

\subsection{Scope}

This paper contributes DAI-specific event construction, a common eligible-pair denominator, a candidate-invariant deadline and an exact condition for one failure mode, with synthetic verification. Its probability, misclassification and accounting tools are established; their alignment to DAI events is new. How often these failures occur in a given service is a question for instrumented data.

SIM1 is an offline verifier with no media player, ad server or audience. Its policies are trigger rules, not SSAI or SGAI implementations, so it ranks neither architecture; its outcomes follow from the event definitions, and its M0 agreement with Proposition~1 checks the generator's algebra, not field behavior. Its path mixtures, latency laws, lifetimes and monetary coefficients are uncalibrated stress inputs. It holds paths and latency draws common to both policies to isolate the readiness mechanism from queueing feedback and policy-dependent abandonment; field collectability, load feedback and policy-dependent reach are open problems (Section VII). Trigger placement and opportunity position are fixed, the grid leaves out recurrent viewing, long-run retention, cross-session caps and bidding equilibrium, and the clock-translated copies test coordinate invariance only. The loss masks model neither correlated outages, invalid-traffic filtering nor uncertainty about which event classes were captured.

A deployment needs a dated, rights-cleared client--server join on its target runtimes that demonstrates capture coverage, the alignment bound \(\varepsilon\) of Section III-D and the common readiness criterion with client evidence (nominal server timestamps cannot stand in for client validity transitions), plus contractual reconciliation of pod value and sensitivity analysis for disputed shared costs (Table~\ref{tab:observability}, Section V-D).

\subsection{Implications for architecture analysis}

An earlier trigger can enlarge the readiness window and expose more nonreaching pairs to work. It can also allow a prepared state to expire before the deadline. The resulting contribution depends on actual events, value, cancellation, reuse and shared costs. Guided insertion can change cacheability, scheduling and observability; stitched insertion can also preload. The documented guided-prefetch heartbeat is one implementation-specific example \cite{r9}.

In synchronized live delivery, many sessions can reach the same ad break within seconds. Concentrated requests can overload the ad decision service; AWS documents retrieval-window and transactions-per-second shaping to limit timeouts and low fill \cite{r8}. Measured field effects enter candidate-specific latency and readiness, conditional pod values \(v_a^{\mathrm{ready}}\) and \(v_a^{\mathrm{miss}}\), variable costs, and reserved-capacity cost \(K_a(N,z)\). Concurrency, timeout and load-control settings belong in the candidate definition and comparison record.

Request-time or late-binding assembly is another candidate. With fixed job-duration and lifetime laws, delaying resolution shortens \(W_{ai}\), reducing expiry exposure while increasing deadline-miss exposure; synchronized request-time work can also concentrate load. Measurement and Equation~\eqref{eq:main-10} settle which effect dominates; under Proposition~1's assumptions, Equation~\eqref{eq:main-12} gives it in closed form.

Common reach and equal coefficients are assumptions to test. When ad work can affect primary playback, retain candidate-specific reach. Keep pod-horizon contribution separate from long-run engagement or retention.

\subsection{Joint uncertainty and decisions}

A difference smaller than its own uncertainty should not decide an architecture. Use the common-cohort contribution contrast defined in Equation~\eqref{eq:main-11}.

Estimate or bound \(\Delta m_{a,b}(N,z)\) jointly, retaining covariance only where assignment and data construction supply it. Genuine pairing can come from randomized paired observations or a non-serving shadow calculation on the same opportunity, not from alternatives seen on separate pairs.

For deterministic sensitivity analysis, let \(\theta\) be the vector of uncertain inputs, let \(\mathcal U\) be a specified nonempty set of jointly admissible values, and write \(\Delta m_{a,b}(N,z;\theta)\) for the difference produced by one allowed input vector. Each evaluated input must satisfy the accounting assumptions and finite-moment conditions. Holding \(N\) and \(z\) fixed, candidate \(a\) is \textbf{robustly dominant} only if \(\inf_{\theta\in\mathcal U}\Delta m_{a,b}(N,z;\theta)>0\), and candidate \(b\) is robustly dominant only if \(\sup_{\theta\in\mathcal U}\Delta m_{a,b}(N,z;\theta)<0\).

For statistical inference, construct a confidence interval directly for \(\Delta m_{a,b}(N,z)\), or map a simultaneous joint confidence region through Equation~\eqref{eq:main-11}. Two marginal intervals do not automatically deliver their individual coverage jointly. A positive lower confidence limit supports a positive difference at the stated confidence level without establishing deterministic robust dominance. Under either analysis, a range containing zero is indeterminate.

The same holds for an externally justified readiness threshold: a point estimate above it is not enough if the one-sided confidence bound falls below it.

\section{Conclusion}

DAI comparisons need separate trigger, reach, state-validity and commercial records on one eligible-pair denominator, scored against a deadline no candidate can move. Under its assumptions, Proposition~1 makes the trade between early and late triggering exact: it states when completion scoring reverses a comparison and shows that such scoring can only shift it toward the earlier trigger. SIM1 reproduces its condition on monotone paths (a check of its generator, not of the field), finds that playhead misclassification of up to 41\% of reach indicators reverses no comparison in the study, and finds completion-scoring reversals in three of nine short-lifetime conditions, all at the one grid latency inside the M0 reversal bands (Section V-C3). Three problems remain open.
\begin{enumerate}
\tightlist
\item \textbf{Collectability.} An instrumented server-guided testbed with a scripted client, such as DASH-IF Livesim2 \cite{r33}, would collect the client transitions behind \(Z_{ai}(t)\) and measure \(\varepsilon\) on one path instead of stipulating it.
\item \textbf{Load feedback.} At a synchronized live break, queueing ties latency to trigger timing; extending Proposition~1 to a policy-dependent latency law remains open.
\item \textbf{Policy-dependent reach.} Candidate-specific reach lies outside Proposition~1; the all-pair contrast stays identified under randomization, but reach-conditioned contrasts need assumptions about pairs that would reach under both \cite{r37}.
\end{enumerate}

\section*{Acknowledgment}

The authors used ChatGPT (OpenAI) and Claude (Anthropic) for drafting and LaTeX preparation throughout the article and supplement, and to check derivations in Sections III--IV and Supplement S1--S5, S7 and S9. ChatGPT was also used for figure, simulation and verification code. The mathematical development, event definitions, analyses and conclusions are the authors' own. The authors reviewed all AI-assisted text and code and take full responsibility for the work.

\textbf{Code and data availability.} Repository: \url{https://github.com/methodtrace/2026-dai-timing}. Release tag: v1.0.0. Zenodo version DOI: 10.5281/zenodo.22774143. Code: Apache-2.0; data, tables, fixtures and the event-dictionary template: CC BY 4.0. The artifact includes source, configuration, 90 seeds and 90 event shards, masks, all tables, 19 fixtures, separate verifiers, file/content hashes, a transport-equivalence audit, the correctly rounded math module with its tests, and an execution history. The arrival diagnostic adds two scoring/checking scripts, its frozen protocol and input manifest, 72 condition rows, 360 replicate rows, 60,000 pair scores, selected event traces and eleven separate known-answer checks; its 60 input shards reuse the same stored data. A separate exact revaluation check preserves both scoring conventions. Section V-C5 reports the verification.

\ifdefined\biobudget\par\vspace*{0.8\textheight}\mbox{}\fi

\begin{thebibliography}{99}
\bibitem{r8} Amazon Web Services, ``How prefetching works,'' \emph{AWS Elemental MediaTailor User Guide}. Accessed: Sep.~9, 2026. {[}Online{]}. Available: \url{https://docs.aws.amazon.com/mediatailor/latest/ug/understanding-prefetching.html}

\bibitem{r11} IAB Technology Laboratory, ``Digital Video Ad Serving Template (VAST).'' Accessed: Sep.~8, 2026. {[}Online{]}. Available: \url{https://iabtechlab.com/standards/vast/}

\bibitem{r12} IAB Technology Laboratory, ``Open Measurement SDK.'' Accessed: Sep.~8, 2026. {[}Online{]}. Available: \url{https://iabtechlab.com/standards/open-measurement-sdk/}

\bibitem{r13} IAB Technology Laboratory, ``Programmatic guide---connected TV with highlights for SSAI.'' Accessed: Sep.~8, 2026. {[}Online{]}. Available: \url{https://iabtechlab.com/standards/ctv-programmatic-guide/}

\bibitem{r14} IAB Technology Laboratory, ``OpenRTB 2.x implementation notes.'' Accessed: Sep.~8, 2026. {[}Online{]}. Available: \url{https://github.com/InteractiveAdvertisingBureau/openrtb2.x/blob/main/implementation.md}

\bibitem{r1} A. Giladi, N. Levy, and T. Griebel, ``Server-guided ad insertion in 6th edition of MPEG DASH,'' in \emph{Proceedings of the 5th Mile-High Video Conference}, pp.~59--65, 2026, doi: 10.1145/3789239.3793274.

\bibitem{r2} R. Pantos, ``HTTP Live Streaming 2nd edition,'' Internet-Draft draft-pantos-hls-rfc8216bis-22, May 2026, work in progress. Accessed: Sep.~8, 2026. {[}Online{]}. Available: \url{https://datatracker.ietf.org/doc/html/draft-pantos-hls-rfc8216bis-22}

\bibitem{r3} DASH Industry Forum, ``Livesim2 URL generator: server-guided ad insertion,'' 2026. Accessed: Sep.~8, 2026. {[}Online{]}. Available: \url{https://livesim2.dashif.org/urlgen/create}

\bibitem{r4} MPEG Group, ``DASH sixth-edition MPD schema,'' branch \texttt{6th-Ed}. Accessed: Sep.~8, 2026. {[}Online{]}. Available: \url{https://github.com/MPEGGroup/DASHSchema/blob/6th-Ed/DASH-MPD.xsd}

\bibitem{r5} International Organization for Standardization and International Electrotechnical Commission, ``ISO/IEC 23009-1:2026: information technology---Dynamic Adaptive Streaming over HTTP (DASH)---part 1: media presentation description and segment formats,'' 6th ed., July 2026. Accessed: Sep.~8, 2026. {[}Online{]}. Available: \url{https://www.iso.org/standard/23009-1}

\bibitem{r6} Apple Inc., ``Explore dynamic pre-rolls and mid-rolls in HLS,'' WWDC21, session 10140, 2021. {[}Online{]}. Available: \url{https://developer.apple.com/videos/play/wwdc2021/10140/}

\bibitem{r7} Amazon Web Services, ``MediaTailor server-guided ad insertion overview and implementation,'' \emph{AWS Elemental MediaTailor User Guide}. Accessed: Sep.~8, 2026. {[}Online{]}. Available: \url{https://docs.aws.amazon.com/mediatailor/latest/ug/server-guided.html}

\bibitem{r9} Amazon Web Services, ``Guided prefetch with manifest heartbeating,'' \emph{AWS Elemental MediaTailor User Guide}. Accessed: Sep.~9, 2026. {[}Online{]}. Available: \url{https://docs.aws.amazon.com/mediatailor/latest/ug/sgai-guided-prefetch.html}

\bibitem{r10} Amazon Web Services, ``MediaTailor server-guided ad insertion feature compatibility matrix,'' \emph{AWS Elemental MediaTailor User Guide}. Accessed: Sep.~9, 2026. {[}Online{]}. Available: \url{https://docs.aws.amazon.com/mediatailor/latest/ug/sgai-feature-compatibility.html}

\bibitem{r15} Google, ``Server guided DAI,'' Google Ad Manager documentation, updated Mar.~3, 2026. Accessed: Sep.~8, 2026. {[}Online{]}. Available: \url{https://developers.google.com/ad-manager/dynamic-ad-insertion/server-guided}

\bibitem{r33} DASH Industry Forum, ``Livesim2,'' repository README, ``Alternative MPD events and SGAI.'' Accessed: Sep.~8, 2026. {[}Online{]}. Available: \url{https://github.com/Dash-Industry-Forum/livesim2}

\bibitem{r16} DASH Industry Forum, ``DASH-IF interoperability points; part 5: ad insertion,'' version 5.0.0, Nov.~2021. {[}Online{]}. Available: \url{https://dashif.org/docs/IOP-Guidelines/DASH-IF-IOP-Part5-v5.0.0.pdf}

\bibitem{r17} Amazon Web Services, ``Support for HLS interstitials in AWS Elemental MediaTailor,'' \emph{AWS for M\&E Blog}. Accessed: Sep.~8, 2026. {[}Online{]}. Available: \url{https://aws.amazon.com/blogs/media/support-for-hls-interstitials-in-aws-elemental-mediatailor/}

\bibitem{r27} Society of Cable Telecommunications Engineers, ``ANSI/SCTE 35-1 2023r2: digital program insertion cueing message---part 1: legacy splice-based and time-based signaling,'' standards catalog. Accessed: Sep.~7, 2026. Catalog scope consulted. {[}Online{]}. Available: \url{https://account.scte.org/standards/library/catalog/scte-35-1-digital-program-insertion-cueing-message-part-1-legacy-splice-based-and-time-based-signaling/}

\bibitem{r28} Society of Cable Telecommunications Engineers, ``ANSI/SCTE 224 2021: event scheduling and notification interface,'' standards catalog. Accessed: Sep.~7, 2026. Catalog scope consulted. {[}Online{]}. Available: \url{https://account.scte.org/standards/library/catalog/scte-224-event-scheduling-and-notification-interface/}

\bibitem{r18} IAB Technology Laboratory, ``OpenRTB 2.6 specification.'' Accessed: Sep.~8, 2026. {[}Online{]}. Available: \url{https://github.com/InteractiveAdvertisingBureau/openrtb2.x/blob/main/2.6.md}

\bibitem{r29} Interactive Advertising Bureau and Media Rating Council, ``Digital video impression measurement guidelines,'' June 2018, sec.~3 and sec.~4.1, pp.~5--8. Accessed: Sep.~8, 2026. {[}Online{]}. Available: \url{https://mediaratingcouncil.org/sites/default/files/Standards/Digital\%20Video\%20Served\%20Impression\%20Measurement\%20Guidelines\%20(MMTF\%20June\%202018).pdf}

\bibitem{r30} Media Rating Council, ``Server-side ad insertion and OTT guidance,'' Aug.~2021, secs. 2--4, especially pp.~7 and 12. Accessed: Sep.~8, 2026. {[}Online{]}. Available: \url{https://mediaratingcouncil.org/sites/default/files/Standards/083021\%20SSAI\%20and\%20OTT\%20Guidance\%20\%20FINAL.pdf}

\bibitem{r31} Amazon Web Services, ``Client-side ad tracking,'' \emph{AWS Elemental MediaTailor User Guide}, ``Client-side reporting workflow'' and ``Server-guided ad insertion.'' Accessed: Sep.~8, 2026. {[}Online{]}. Available: \url{https://docs.aws.amazon.com/mediatailor/latest/ug/ad-reporting-client-side.html}

\bibitem{r19} V. N. Padmanabhan and J. C. Mogul, ``Using predictive prefetching to improve World Wide Web latency,'' \emph{ACM SIGCOMM Computer Communication Review}, vol.~26, no. 3, pp.~22--36, 1996, doi: 10.1145/235160.235164.

\bibitem{r20} M. Crovella and P. Barford, ``The network effects of prefetching,'' in \emph{Proceedings of IEEE INFOCOM '98}, vol.~3, pp.~1232--1239, 1998, doi: 10.1109/INFCOM.1998.662937.

\bibitem{r21} C. L. Liu and J. W. Layland, ``Scheduling algorithms for multiprogramming in a hard-real-time environment,'' \emph{Journal of the ACM}, vol.~20, no. 1, pp.~46--61, 1973, doi: 10.1145/321738.321743.

\bibitem{r22} R. Dubin, A. Dvir, O. Hadar, T. Frid, and A. Vesker, ``Novel ad insertion technique for MPEG-DASH,'' in \emph{2015 IEEE 12th Consumer Communications and Networking Conference}, pp.~582--587, 2015, doi: 10.1109/CCNC.2015.7158038.

\bibitem{r23} N. Le Scouarnec, C. Neumann, and G. Straub, ``Cache policies for cloud-based systems: to keep or not to keep,'' in \emph{2014 IEEE 7th International Conference on Cloud Computing}, pp.~1--8, 2014, doi: 10.1109/CLOUD.2014.11.

\bibitem{r24} S. Shunmuga Krishnan and R. K. Sitaraman, ``Video stream quality impacts viewer behavior: inferring causality using quasi-experimental designs,'' in \emph{Proceedings of the ACM Internet Measurement Conference}, pp.~211--224, 2012, doi: 10.1145/2398776.2398799.

\bibitem{r25} M. Plakia et al., ``Should I stay or should I go: analysis of the impact of application QoS on user engagement in YouTube,'' \emph{ACM Transactions on Modeling and Performance Evaluation of Computing Systems}, vol.~5, no. 2, article 9, pp.~1--32, 2020, doi: 10.1145/3377873.

\bibitem{r37} C. E. Frangakis and D. B. Rubin, ``Principal stratification in causal inference,'' \emph{Biometrics}, vol.~58, no. 1, pp.~21--29, 2002, doi: 10.1111/j.0006-341X.2002.00021.x.

\bibitem{r34} P. Glasserman and D. D. Yao, ``Some guidelines and guarantees for common random numbers,'' \emph{Management Science}, vol.~38, no. 6, pp.~884--908, 1992, doi: 10.1287/mnsc.38.6.884.

\bibitem{r36} C. F. Manski, \emph{Partial Identification of Probability Distributions}. Springer, 2003, doi: 10.1007/b97478.

\bibitem{r38} K. T. Copeland, H. Checkoway, A. J. McMichael, and R. H. Holbrook, ``Bias due to misclassification in the estimation of relative risk,'' \emph{American Journal of Epidemiology}, vol.~105, no. 5, pp.~488--495, 1977, doi: 10.1093/oxfordjournals.aje.a112408.

\bibitem{r39} J. J. Yland, A. K. Wesselink, T. L. Lash, and M. P. Fox, ``Misconceptions about the direction of bias from nondifferential misclassification,'' \emph{American Journal of Epidemiology}, vol.~191, no. 8, pp.~1485--1495, 2022, doi: 10.1093/aje/kwac035; correction, vol.~191, no. 12, p.~2123, doi: 10.1093/aje/kwac129.

\bibitem{r32} Amazon Web Services, ``AWS Elemental MediaTailor manifest logs description and event types,'' \emph{AWS Elemental MediaTailor User Guide}. Accessed: Sep.~8, 2026. {[}Online{]}. Available: \url{https://docs.aws.amazon.com/mediatailor/latest/ug/log-types.html}

\bibitem{r26} P. C. Sruthi, S. Rao, and B. Ribeiro, ``Pitfalls of data-driven networking: a case study of latent causal confounders in video streaming,'' in \emph{Proceedings of the 2020 Workshop on Network Meets AI \& ML}, pp.~42--47, 2020, doi: 10.1145/3405671.3405815.

\bibitem{r35} L. D. Brown, T. T. Cai, and A. DasGupta, ``Interval estimation for a binomial proportion,'' \emph{Statistical Science}, vol.~16, no. 2, pp.~101--117, 2001, doi: 10.1214/ss/1009213286.
\end{thebibliography}
\end{document}